\documentclass[twoside,twocolumn,9pt]{article}
\usepackage{extsizes}
\usepackage[super,sort&compress,comma]{natbib} 
\usepackage[version=3]{mhchem}
\usepackage[left=1.5cm, right=1.5cm, top=1.785cm, bottom=2.0cm]{geometry}
\usepackage{balance}
\usepackage{mathptmx}
\usepackage{sectsty}
\usepackage{graphicx} 
\usepackage{lastpage}
\usepackage[format=plain,justification=justified,singlelinecheck=false,font={stretch=1.125,small,sf},labelfont=bf,labelsep=space]{caption}
\usepackage{float}
\usepackage{fancyhdr}
\usepackage{fnpos}
\usepackage[english]{babel}
\addto{\captionsenglish}{%
  
}
\usepackage{array}
\usepackage{droidsans}
\usepackage{charter}
\usepackage[T1]{fontenc}
\usepackage[usenames,dvipsnames]{xcolor}
\usepackage{setspace}
\usepackage[compact]{titlesec}
\usepackage{hyperref}

\usepackage{epstopdf}

\definecolor{cream}{RGB}{222,217,201}

\begin{document}

\pagestyle{fancy}
\thispagestyle{plain}
\fancypagestyle{plain}{
\renewcommand{\headrulewidth}{0pt}
}

\makeFNbottom
\makeatletter
\renewcommand\LARGE{\@setfontsize\LARGE{15pt}{17}}
\renewcommand\Large{\@setfontsize\Large{12pt}{14}}
\renewcommand\large{\@setfontsize\large{10pt}{12}}
\renewcommand\footnotesize{\@setfontsize\footnotesize{7pt}{10}}
\makeatother

\renewcommand{\thefootnote}{\fnsymbol{footnote}}
\renewcommand\footnoterule{\vspace*{1pt}%
\color{cream}\hrule width 3.5in height 0.4pt \color{black}\vspace*{5pt}} 
\setcounter{secnumdepth}{5}

\makeatletter 
\renewcommand\@biblabel[1]{#1}            
\renewcommand\@makefntext[1]%
{\noindent\makebox[0pt][r]{\@thefnmark\,}#1}
\makeatother 
\renewcommand{\figurename}{\small{Fig.}~}
\sectionfont{\sffamily\Large}
\subsectionfont{\normalsize}
\subsubsectionfont{\bf}
\setstretch{1.125} 
\setlength{\skip\footins}{0.8cm}
\setlength{\footnotesep}{0.25cm}
\setlength{\jot}{10pt}
\titlespacing*{\section}{0pt}{4pt}{4pt}
\titlespacing*{\subsection}{0pt}{15pt}{1pt}

\fancyfoot{}
\fancyfoot[LO,RE]{\vspace{-7.1pt}\includegraphics[height=9pt]{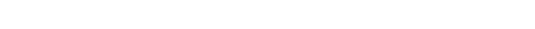}}
\fancyfoot[CO]{\vspace{-7.1pt}\hspace{11.9cm}\includegraphics{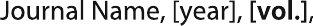}}
\fancyfoot[CE]{\vspace{-7.2pt}\hspace{-13.2cm}\includegraphics{head_foot/RF}}
\fancyfoot[RO]{\footnotesize{\sffamily{1--\pageref{LastPage} ~\textbar  \hspace{2pt}\thepage}}}
\fancyfoot[LE]{\footnotesize{\sffamily{\thepage~\textbar\hspace{4.65cm} 1--\pageref{LastPage}}}}
\fancyhead{}
\renewcommand{\headrulewidth}{0pt} 
\renewcommand{\footrulewidth}{0pt}
\setlength{\arrayrulewidth}{1pt}
\setlength{\columnsep}{6.5mm}
\setlength\bibsep{1pt}

\makeatletter 
\newlength{\figrulesep} 
\setlength{\figrulesep}{0.5\textfloatsep} 

\newcommand{\topfigrule}{\vspace*{-1pt}%
\noindent{\color{cream}\rule[-\figrulesep]{\columnwidth}{1.5pt}} }

\newcommand{\botfigrule}{\vspace*{-2pt}%
\noindent{\color{cream}\rule[\figrulesep]{\columnwidth}{1.5pt}} }

\newcommand{\dblfigrule}{\vspace*{-1pt}%
\noindent{\color{cream}\rule[-\figrulesep]{\textwidth}{1.5pt}} }

\makeatother

\newcommand{\ket}[1]{\ensuremath{\left|#1\right\rangle}}
\newcommand{\bra}[1]{\ensuremath{\langle#1|}}

\twocolumn[
  \begin{@twocolumnfalse}
{\includegraphics[height=30pt]{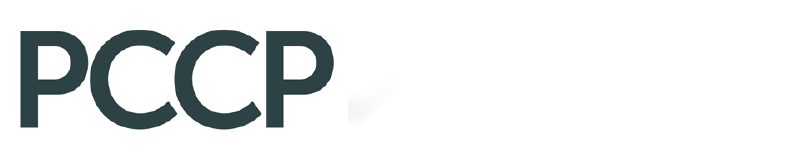}\hfill\raisebox{0pt}[0pt][0pt]{\includegraphics[height=55pt]{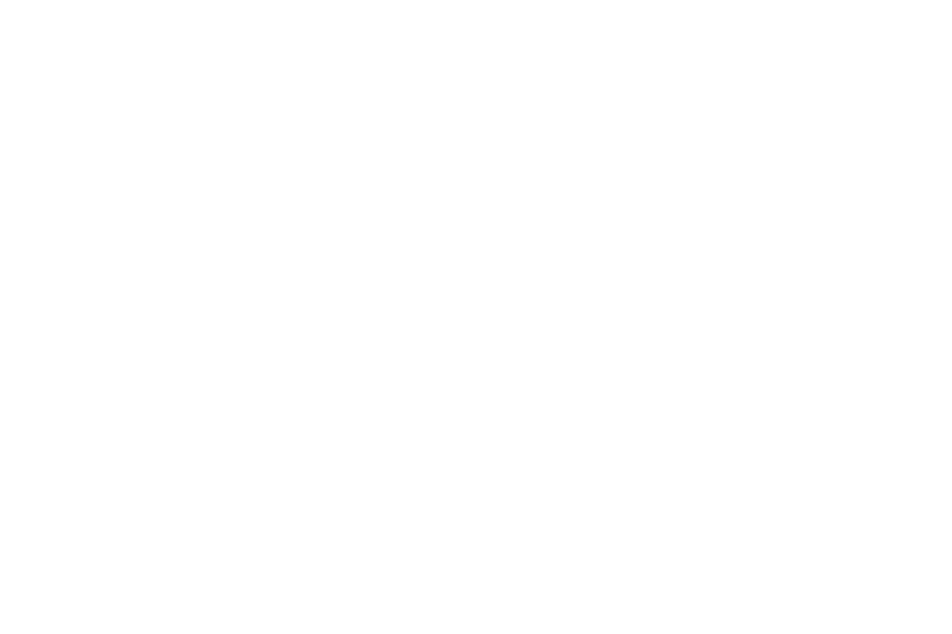}}\\[1ex]
\includegraphics[width=18.5cm]{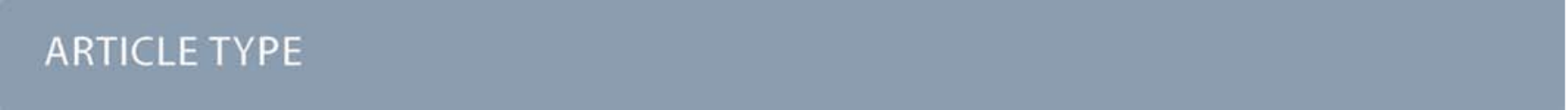}}\par
\vspace{1em}
\sffamily
\begin{tabular}{m{4.5cm} p{13.5cm} }

\includegraphics{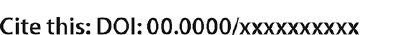} & \noindent\LARGE{\textbf{Electrically tuned hyperfine spectrum in neutral Tb(II)(Cp$^{\rm{iPr5}})_2$ single-molecule magnet$^\dag$}} \\
\vspace{0.3cm} & \vspace{0.3cm} \\

& \noindent\large{Robert L. Smith,\textit{$^{a}$} Aleksander L. Wysocki,\textit{$^{b}$} and Kyungwha Park\textit{$^{b}$}} \\

\includegraphics{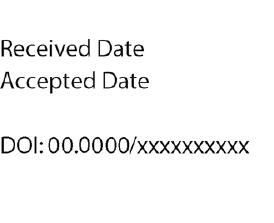} & \noindent\normalsize{
Molecular spin qubits with long spin coherence time as well as non-invasive operation methods on such qubits are in high demand. It was shown that both molecular electronic and nuclear spin levels can be used as qubits. In solid state systems with dopants, an electric field was shown to effectively change the spacing between the nuclear spin qubit levels when the electron spin density is high at the nucleus of the dopant. Inspired by such solid-state systems, we propose that divalent lanthanide (Ln) complexes with an unusual electronic configuration of Ln$^{2+}$ have a strong interaction between the Ln nuclear spin and the electronic degrees of freedom, which renders electrical tuning of the interaction. As an example, we study electronic structure and hyperfine interaction of the $^{159}$Tb nucleus in a neutral Tb(II)(Cp$^{\rm{iPr5}}$)$_2$ single-molecule magnet (SMM), which exhibits unusually long magnetization relaxation time, using the complete active space self-consistent field (CASSCF) method with spin-orbit interaction included within the restricted active space state interaction (RASSI). Our calculations show that the low-energy states arise from $4f^8(6s,5d_{z^2})^1$, $4f^8(5d_{x^2-y^2})^1$, and $4f^8(5d_{xy})^1$ configurations. We compute the hyperfine interaction parameters and the electronic-nuclear spectrum within our multiconfigurational approach. We find that the hyperfine interaction is about one order of magnitude greater than that for Tb(III)Pc$_2$ SMMs. This stems from the strong Fermi contact interaction between the Tb nuclear spin and the electron spin density at the nucleus that originates from the occupation of the $(6s,5d)$ orbitals. We also uncover that the response of the Fermi contact term to electric field results in electrical tuning of the electronic-nuclear level separations. This hyperfine Stark effect may be useful for applications of molecular nuclear spins for quantum computing.
} \\

\end{tabular}

 \end{@twocolumnfalse} \vspace{0.6cm}

  ]

\renewcommand*\rmdefault{bch}\normalfont\upshape
\rmfamily
\section*{}
\vspace{-1cm}


\footnotetext{\textit{$^{a}$~Department of Chemistry, Virginia Tech, Blacksburg, Virginia 24061, USA.}}
\footnotetext{\textit{$^{b}$~Department of Physics, Virginia Tech, Blacksburg, Virginia 24061, USA. Fax: 1 540 231 7511; Tel: 1 540 231 5533; E-mail: kyungwha@vt.edu; alexwysocki2@gmail.com }}

\footnotetext{\dag~Electronic Supplementary Information (ESI) available: Relative energies of the high-spin and low-spin spin-free states from CASSCF(13,14) calculations; relative energies of the several tens of the CASSCF(13,14)-SO-RASSI high-spin states.]. See DOI: 00.0000/00000000.}

\section{Introduction}
A variety of solid-state systems have been proposed and used for quantum computing applications. The experimental endeavor of using molecules for such applications is fairly nascent,\cite{Wernsdorfer2019,Atzori2018,Gaita2019} although a first theoretical proposal of such an idea dates back to almost twenty years ago.\cite{Leuenberger2001} So far, the majority of effort has been focused on using either molecular electronic spins or nuclear spins. Both directions have its own merits and drawbacks. The first approach is easier and faster to operate but with shorter spin coherence time, while the latter is slower to operate but with longer coherence time. There are extensive studies of hybrid systems where the merits of both approaches are taken into account in solid-state systems.\cite{Awschalom2018}

Lanthanide-based molecules have properties useful for quantum computing applications\cite{Thiele2014,Shiddiq2016,Godfrin2017,Pineda2018,Wernsdorfer2019,Atzori2018,Gaita2019} such as strong spin-orbit (SO) interaction and strong interaction between the lanthanide nuclear spin and the electronic degrees of freedom, i.e. hyperfine interaction. Properties of the molecules can be tailored by varying the lanthanide element, ligand or oxidation state, or by a judicious choice of external perturbation. Terbium (Tb) based single-molecule magnets (SMMs) such as Tb(III)Pc$_2$ (Pc=pthalocyanine)\cite{Ishikawa2003} were reported to remain stable within single-molecule transistor set-ups\cite{Thiele2013} and to exhibit strong hyperfine interaction between the $^{159}$Tb (100\% natural abundance) nuclear spin ($I=3/2)$ and the magnetic moment of the electron.\cite{Ishikawa2005} Tb(III)Pc$_2$ molecules were also shown to reveal significant modulation of the hyperfine interaction with external electric field.\cite{Thiele2014,Godfrin2017,Pineda2018,Wernsdorfer2019} Such a hyperfine Stark effect is a molecular manifestation of the concept proposed by Kane\cite{Kane1998} on phosphorus (P) dopants in silicon (Si) solids. In this proposal, delocalized electron spin density of the P dopant at $^{31}$P nucleus results in hyperfine interaction induced by the Fermi contact (FC) term that can be greatly modulated by applying voltage or an electric field. This possibility is very appealing since logical operations for quantum computing can be manipulated by electric field\cite{Kane1998} rather than magnetic field. Note that electric field can be applied locally, whereas that is challenging for magnetic field. In the case of Tb-based SMMs realization and manipulation of molecular nuclear spin qubits were also facilitated by using the significant hyperfine Stark effect.

\begin{figure}[h]
\centering
\includegraphics[height=5cm]{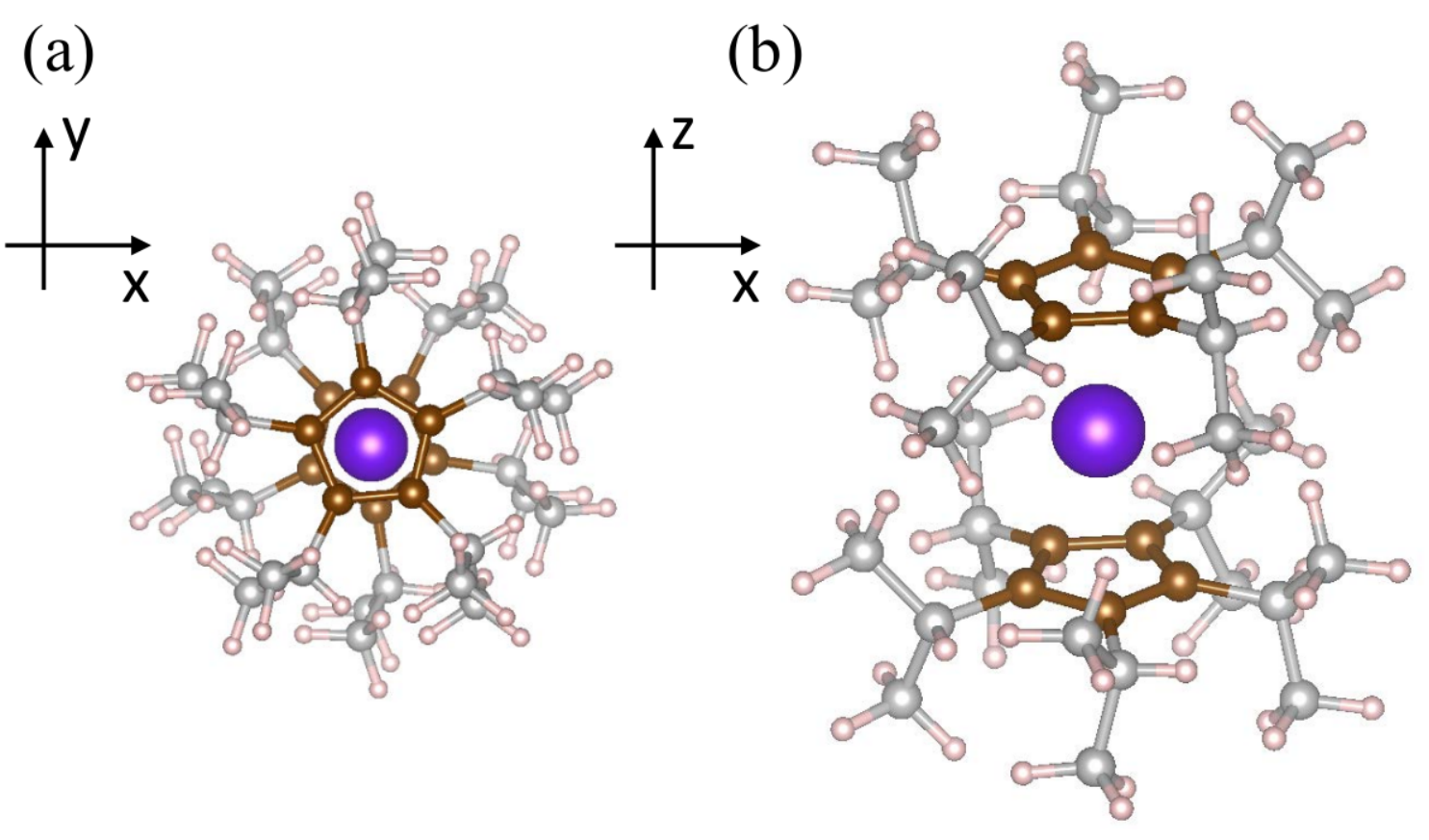}
\caption{(a) Top view and (b) side view of experimental\cite{Gould2019} geometry of the neutral Tb(II)(Cp$^{\rm{iPr5}}$)$_2$ molecule. A color scheme is as follows: Tb (purple), C in the Cp rings (brown), the rest of C (gray), H (pale pink). The symmetry of the molecule is slightly deviated from $D_{5d}$ point group. The magnetic easy axis coincides with the $z$ axis.}
\label{fig:geo}
\end{figure}

Despite the significant hyperfine Stark effect, Tb(III)Pc$_2$ SMMs have a very small FC term due to negligible electron spin density at the Tb nucleus. Keeping in mind that in the original proposal by Kane,\cite{Kane1998} a large FC term is a key element in the strong hyperfine Stark effect, we search for magnetic molecules with a large FC term in the hope for a strong hyperfine Stark effect. One class of molecules that possibly fit into this criterion are lanthanide(II)-based molecules. Divalent lanthanide complexes have been known to be rarely synthesized or unstable at ambient environment,\cite{MacDonald2013} compared to trivalent lanthanide complexes.\cite{Woodruff2013} Recently, several divalent lanthanide-based molecules have been synthesized in a form of crystals with stability at room temperature.\cite{MacDonald2012,Gould2019,Fieser2015,Evans2016,Fieser2017,Huh2018,Ryan2018,Meihaus2015} Magnetic measurements suggest that the stable electronic configuration of divalent lanthanide ions (with $n+1$ valence electrons) are likely $4f^n(6s,5d)^1$ rather than $4f^{n+1}$, where $(6s,5d)$ denotes strong hybridization between $6s$ and $5d$ orbitals. One of such molecules is a neutral Tb(II)(Cp$^{\rm{iPr5}}$)$_2$ SMM (Cp=pentaisopropylcyclopentadienyl) which exhibits unusually long magnetization relaxation time ($\sim$10$^3$ s) and magnetic hysteresis until 55~K.\cite{Gould2019} As shown in Fig.~\ref{fig:geo}, the molecule consists of two pentagon-shaped Cp rings above and below the divalent Tb ion which has approximate $D_{5d}$ point group symmetry. Magnetic susceptibility measurement supports the idea that the Tb$^{2+}$ ion has a stable electronic configuration of $4f^8(6s,5d)^1$ rather than $4f^9$.\cite{Gould2019} Electronic structure and magnetic properties of this compound were studied using density-functional theory (DFT) calculations.\cite{Gould2019} However, nearly degenerate $4f$ orbitals demand theoretical treatment beyond DFT. So far, multiconfigurational or multireference studies of this compound have not been done. Overall, {\it ab-initio} studies (beyond DFT) of divalent lanthanide complexes are scarce.\cite{Zhang2020} Hyperfine interaction of this compound has not been examined before.

Here we uncover the nature of the hyperfine interaction of $^{159}$Tb nucleus in the neutral Tb(II)(Cp$^{\rm{iPr5}}$)$_2$ SMM using the complete active space self-consistent field (CASSCF) method with SO interaction included within the restricted active space state interaction (RASSI). We first identify the electronic structure of the ground state and low-lying excited states using the CASSCF method with SO-RASSI, and then calculate the hyperfine interaction of the Tb nucleus projected onto the ground Kramers doublet with and without small external electric field. Our study may shed light into search for molecules with strong hyperfine Stark effect and its applications to control of nuclear spin levels for quantum computing.

\section{Computational details}

\begin{figure*}[h]
\centering
\includegraphics[height=10.cm]{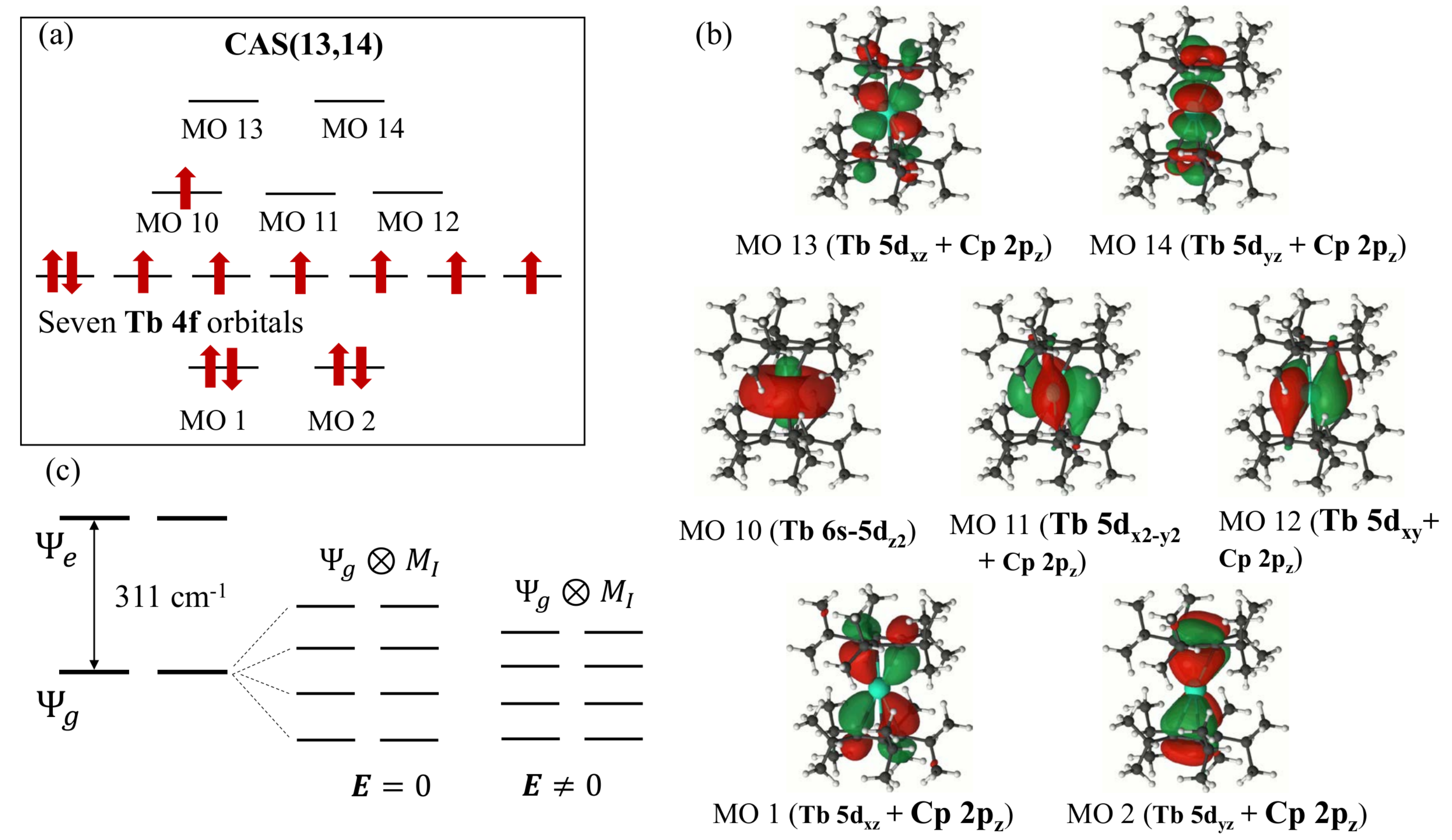}
\caption{(a) Schematic diagram of the active space used in this work, CAS(13,14), where up and down arrows represent $\alpha$ and $\beta$ spin of electron. (b) Active non-$4f$ molecular orbitals (MO) obtained from our CASSCF(13,14) calculation with state average over 21 roots. The two doubly occupied orbitals (MO 1 and 2) are from strong hybridization between the Cp rings 2$p_z$ orbitals and the Tb 5d$_{xz}$ and 5d$_{yz}$ orbitals. In this case, a larger font size indicates a larger weight. MO 3-9 are pure $4f$ orbitals. MO 10 arises from strongly hybridized Tb 6s and 5d$_z^2$ orbitals, while MO 11 and 12 are from mainly Tb 5d$_{x^2-y^2}$ and 5d$_{xy}$ orbitals hybridized with the Cp $2p_z$ orbitals, respectively. Here a larger weight is shown as a larger font size. MO 13 and 14 are from Tb 5d$_{xz}$ and 5d$_{yz}$ hybridized with the Cp ring 2$p_z$ orbitals. The state-average occupation numbers of MO 1-14 are 1.9928, 1.9928, 1.1426, 1.1427, 1.1427, 1.1427, 1.1427, 1.1426, 1.1427, 0.3333, 0.3331, 0.3330, 0.0082, 0.0082, respectively. The visualization is made with isosurface value of 0.03 using LUSCUS program.\cite{LUSCUS} (c) Schematic diagram of level separations with and without electric field (${\bf E}\neq0$ and ${\bf E}=0$). The electronic ground state $\Psi_g$ (Kramers doublet) is separated from the electronic first-excited state $\Psi_e$ (Kramers doublet) by 311~cm$^{-1}$ from our CASSCF-SO-RASSI calculation. Each electronic level is split into four electronic-nuclear levels due to the hyperfine interaction of the $^{159}$Tb nuclear spin ($I=3/2$). Since the electronic separation energy is at least more than two orders of magnitude greater than separation energies of electronic-nuclear levels $\Psi_g \otimes M_I$, we consider only the hyperfine interaction for the electronic ground doublet. The separations of the electronic-nuclear levels can be modified by ${\bf E}$ field.}
\label{fig:CAS}
\end{figure*}

We use the experimental geometry reported in Ref.~\citenum{Gould2019} without further structure relaxation. Our {\it ab-initio} calculations are carried out using the MOLCAS code (version 8.2)\cite{molcas} with the scalar relativistic effect (based on Douglas-Kroll-Hess Hamiltonian\cite{Douglass1974,Hess1986}) using relativistically contracted atomic natural orbitals (ANO-RCC) basis sets: polarized valence triple-$\zeta$ quality (VTZP) for the Tb ion, polarized valence double-$\zeta$ quality (VDZP) for all the C atoms, and valence double-$\zeta$ quality (VDZ) for the H atoms.

The electronic structure is computed in a two-step procedure. First, in the absence of SO interaction, spin-free eigenstates (roots) are found using the state-averaged (SA) CASSCF method.\cite{Roos1980,Siegbahn1981} Second, in the subspace of the spin-free roots, SO interaction is included within the atomic mean-field approximation,\cite{Hess1996} using the restricted active space state interaction (RASSI) method.\cite{rassi} The same two-step procedure as above is used when a homogeneous electric field is applied along the $z$ axis to the molecule.

Starting from the stable electronic configuration of a trivalent Tb ion, i.e. $4f^8$, one can think of $4f^8(6s,5d)^1$ and $4f^9$ as possible configurations of a divalent Tb ion. In the $4f^9$ configuration, there is only one low-energy spin state with total spin $S=5/2$. In the $4f^8(6s,5d)^1$ configuration, the Tb $4f$ spin $S_{4f}=3$ from $4f^8$ can be parallel or antiparallel to the spin $S_{5d}=1/2$ from $(6s,5d)^1$ as long as we are interested in a low-energy spectrum. These cases correspond to two low-energy spin states such as total spin $S=7/2$ and $S=5/2$. The former state is referred to as high spin (HS), while the latter low spin (LS). (The concept of HS and LS in our work qualitatively differs from the usual context of HS and LS in 
spin-crossover molecules.\cite{Nihei2007,Halcrow2011})

In the $4f^9$ configuration, since the Tb $4f$ orbitals are highly localized, nine electrons and seven 4$f$ orbitals can form a reasonable active space. In this case, 21 roots are used in the state average. Using this active space for the $4f^9$ configuration and the optimal active space discussed below for the $4f^8(6s,5d)^1$ configuration, we check that the ground-state energy of the $4f^9$ configuration is $\sim$~4.4 eV higher than that of the $4f^8(6s,5d)^1$ configuration. This result agrees with the previous DFT calculation and the experimental data that rules out $4f^9$ as the stable configuration of the Tb$^{2+}$ ion in the neutral Tb(II)(Cp$^{\rm{iPr5}}$)$_2$ molecule.\cite{Gould2019} Therefore, we henceforth consider only the $4f^8(6s,5d)^1$ configuration.

The choice of the active space is critical in a SA-CASSCF calculation, especially in the $4f^8(6s,5d)^1$ configuration. The active space should include seven $4f$ orbitals as well as five or six of $(5d,6s)$-like orbitals. In addition, the $(5d,6s)$-like orbitals (more delocalized than the $4f$ orbitals) can have significant hybridization with $\pi/\pi^*$ orbitals of the Cp rings. Therefore, some of these ligand orbitals may also need to be included in the active space. Since the size of the active space is practically limited due to high computational cost, we perform extensive tests on different sets of active spaces analyzing state-average and natural occupation numbers of the active orbitals and their effects on the energetics of the system. We find that only five of the $(6s,5d)$-like orbitals are needed to be included in the active space. These include $5d_{x^2-y^2}$, $5d_{xy}$, $5d_{xz}$, $5d_{yz}$-like orbitals as well as $(5d_{z^2},6s)$ hybrid orbital (see Fig.~\ref{fig:CAS}b). Note that the first four of these $5d$-like orbitals show significant hybridization with $2p_z$ orbitals from the Cp rings. Importantly, another $6s$-like orbital (hybridizing strongly with C $2p$ orbitals) does not need to be included in the active space since its natural occupation number is zero (incidentally it is also true for the Tb $6p$ orbitals). We also find that only two of the $\pi/\pi^*$ orbitals are crucial in the active space. These are nominally doubly occupied $\pi$ orbitals from the Cp rings that show strong hybridization with $5d_{xz}$, $5d_{yz}$ orbitals (see Fig.~\ref{fig:CAS}b). Therefore, the optimal active space is CAS(13,14) that consists of 13 active electrons in 14 active orbitals as shown schematically in Fig.~\ref{fig:CAS}a.

The choice of the number of roots to use in the state-average procedure is also non-tirivial in the $4f^8(6s,5d)^1$ configuration. The $4f^8$ configuration suggests to use seven roots. However, our calculations reveal that there is no significant energy gap between 7$\text{th}$ and 8$\text{th}$ roots. Clearly, the fact that there are many $(6s,5d)^1$ configurations cannot be ignored. In fact, we determine that at low energies there are three relevant $(6s,5d)^1$-type configurations: $(5d_{z^2}/6s)^1$, $(5d_{x^2-y^2})^1$, and $(5d_{xy})^1$ (this is discussed in more detail in Sec.~3.1). Indeed, the first 21 roots are grouped together in energy and there is a gap ($\sim$3 eV) between the 21$\text{st}$ and 22$\text{nd}$ roots. Therefore, 21 roots are used in the state-average procedure as well as in the RASSI calculations.

After obtaining the electronic structure, a magnetic hyperfine matrix ${\bf A}$ and nuclear quadrupole tensor ${\bf P}$ are calculated by projecting the microscopic interactions onto the electronic ground Kramers doublet as discussed in detail in Ref.~\citenum{Wysocki2020}. The projection on the ground doublet is fully justified since the excitation energy of the first-exited doublet ($\sim$300~cm$^{-1}$) is much larger than the maximum level splitting of the Tb nuclear levels ($< 1.0$~cm$^{-1}$). Finally, the $\mathbf{g}$ matrix of the ground doublet is calculated using the SINGLE\_ANISO module\cite{chibotaru2012ab} of the MOLCAS code. The similar procedure to above is applied to calculations of the hyperfine interaction parameters in the presence of a homogeneous ${\bf E}$ field applied along the $z$ axis.

\section{Results and Discussion}

We present the electronic energy spectrum obtained from the CASSCF-SO-RASSI calculation. Then we show the hyperfine and nuclear quadrupole interactions projected onto the ground Kramers doublet with and without ${\bf E}$ field. We then construct an effective spin Hamiltonian and examine the electronic-nuclear spectrum. The effects of an external magnetic ${\bf B}$ field and an ${\bf E}$ field on the spectrum are then discussed.

\subsection{Electronic structure}

\begin{figure}[h]
\centering
\includegraphics[height=4.5cm]{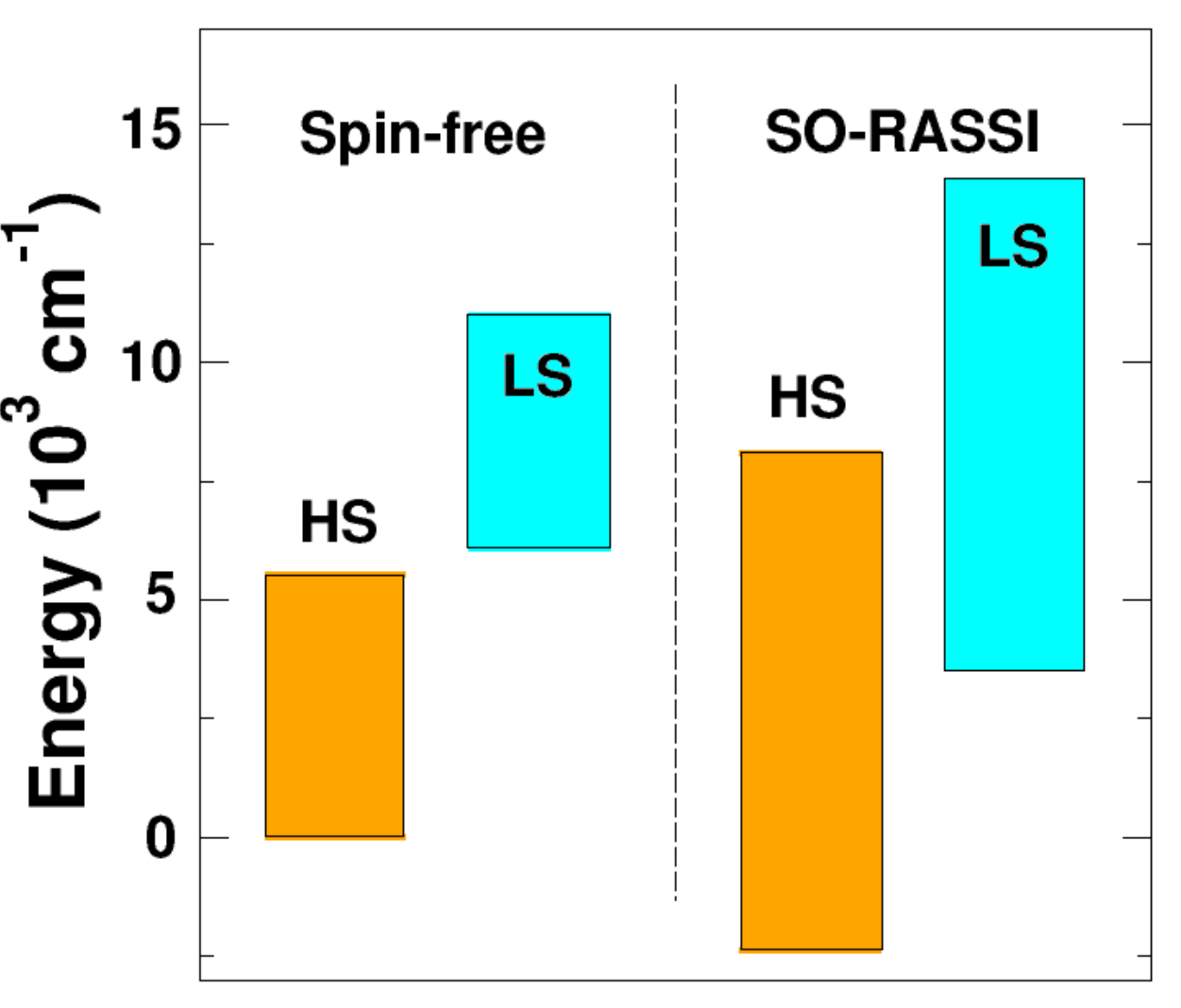}
\caption{The energy spectra of the HS and LS spin-free states and CASSF-SO-RASSI states relative to the ground-state energy of the HS spin-free state. For simplicity, the individual levels are not separately shown and the spectra are shown as rectangular bands that represent the energy range over which the energy levels are distributed.}
\label{fig:comp}
\end{figure}

The CASSCF(13,14) calculations (with state average over 21 roots) show that the lowest HS spin-free energy is 6070~cm$^{-1}$ lower than the lowest LS spin-free energy and that the HS spin-free energy of the 21st root is even lower than the latter energy. (See the ESI$^\dag$ for the HS and LS spin-free energies.) Since the lowest HS spin-free energy is reduced by 2396~cm$^{-1}$ with SO interaction (Fig.~\ref{fig:comp}), the energy difference the lowest HS and LS states is about three times larger than the SO interaction. Previous DFT calculations\cite{Gould2019} show that the HS state is also more stable than the LS state but the energy difference between them is much smaller than our result. Considering our analysis, only the HS state is relevant to the low-energy spectrum, and so we henceforth discuss only the HS state.


In order to have insight into the low-energy electronic energy spectrum, we characterize the four lowest energy spin-free states (or roots). The first and second spin-free states are primarily relevant to the hyperfine interaction that is discussed in Sec.~3.2. They consist of mainly $4f^8(6s,5d_{z^2})^1$ with a small weight from $4f^8(5d_{x^2-y^2},5d_{xy})^1$ (Table~\ref{tab:SF}). The third and fourth spin-free states comprise almost equal weights from $4f^8(5d_{x^2-y^2},5d_{xy})^1$ and $4f^8(6s,5d_{z^2})^1$. This analysis suggests that the exchange coupling mechanism within the Tb$^{2+}$ ion is more complex than the single-electron picture\cite{Anderson1989} relying on a 100\% contribution of $4f^8(6s,5d_{z^2})^1$ configuration. In this work, however, we do not discuss the mechanism of the exchange coupling since the focus is on the hyperfine interaction of the ground Kramers doublet.

\begin{table}[h]
\small
\caption{\ Energies and main contributing configurations for the four lowest energy HS ($S=7/2$) spin-free (SF) states calculated from CASSCF(13,14). The energy is relative to the lowest-energy (or first) SF energy. Here only dominant contributions with corresponding weights are listed}
\label{tab:SF}
\begin{tabular*}{0.48\textwidth}{@{\extracolsep{\fill}}lll}
    \hline
    SF state & Energy (cm$^{-1}$) & Configurations (weight) \\
    \hline
    $\Phi_1$ &   0     & $(4f)^8(6s,5d_{z^2})^1$ (78\%) \\
             &         & $(4f)^8(5d_{x^2-y^2},5d_{xy})^1$ (18\%) \\
    $\Phi_2$ &   0.130 & $(4f)^8(6s,5d_{z^2})^1$ (78\%) \\
             &         & $(4f)^8(5d_{x^2-y^2},5d_{xy})^1$ (18\%) \\
    $\Phi_3$ & 765.327 & $(4f)^8(5d_{x^2-y^2},5d_{xy})^1$ (56\%), \\
             &         & $(4f)^8(6s,5d_{z^2})^1$ (40\%) \\
    $\Phi_4$ & 814.517 & $(4f)^8(5d_{x^2-y^2},5d_{xy})^1$ (54\%), \\
             &         & $(4f)^8(6s,5d_{z^2})^1$ (42\%) \\
    \hline
\end{tabular*}
\end{table}

Next we analyze the energies obtained from the CASSCF-SO-RASSI calculations. With SO-RASSI, all energies are doubly degenerate due to Kramers theorem. The excitation energy of the first-excited doublet is quite high like 311~cm$^{-1}$ due to the strong uniaxial magnetic anisotropy along the $z$ axis, and this value is similar to that of Tb(III)Pc$_2$ SMMs.\cite{Wysocki2020} Interestingly, while $J=13/2$ multiplet structure can be expected from the orbital angular momentum $L=3$ and the spin angular momentum $S=7/2$, the energy spectrum is not consistent with such description since there is no large energy gap between the 14$^\text{th}$ and 15$^\text{th}$ levels (Table~S2 in the ESI$^\dag$). This result suggests that the total angular momentum ${\bf J}$ is {\it not} a good quantum number in this system

Let us now examine the characteristics of four low-energy CASSCF-SO-RASSI wave functions. As shown in Table~\ref{tab:rassi}, the ground doublet $\Psi_g$ is mainly a direct product of $|S=7/2, M_s=\pm 7/2 \rangle$ and a linear combination of the first and second spin-free states, where $M_s$ is the eigenvalue of the $z$ component of the spin operator $S_z$, where the magnetic easy axis coincides with the $z$ axis. We compute the diagonal element of the ${\bf g}$ matrix projected onto the ground doublet, finding that the largest diagonal element ($g_{zz}$) is 19.98985 with the other diagonal elements of an order of $10^{-8}$. The calculated $g_{zz}$ value for the doublet is close to the expected value of $2g_{J}J=19.99998$ when one takes the Lande $g$ factor $g_J$=1.53846 for $L=3$, $S=7/2$, and $J=13/2$. The first-excited doublet $\Psi_{e}$ has some contributions from a direct product of $|S=7/2, M_s=\pm 7/2 \rangle$ and the third and fourth spin-free states in addition to the expected major contributions listed in Table~\ref{tab:rassi}. The characteristics of $\Psi_{e}$ also corroborates that ${\bf J}$ is {\it not} a good quantum number.

\begin{table*}[h]
\small
\caption{\ Calculated energies, ${\bf g}$-matrix, and characteristics of the ground and first-excited Kramers doublets $\Psi_{g}$ and $\Psi_{e}$ (Fig.~\ref{fig:CAS}(c)). Here $\Phi_i$ ($i=1,2,3,4$) are the HS spin-free states listed in Table~\ref{tab:SF}. The wave functions are approximate since only dominant contributions are shown}
\label{tab:rassi}
\begin{tabular*}{0.98\textwidth}{@{\extracolsep{\fill}}lllll}
    \hline
    Wave function & Energy (cm$^{-1}$) & $g_{zz}$ & $g_{xx,yy}$ & Characteristics of wave function \\
    \hline
    $\Psi_{g}$    & 0       & 19.98985 & 0.00000 & $\frac{1}{\sqrt{2}}(\Phi_1 + i \Phi_2) | S=7/2, M_s=-7/2 \rangle$ \\
                  &         &          &         & $\frac{1}{\sqrt{2}}(\Phi_1 - i \Phi_2) | S=7/2, M_s=+7/2 \rangle$ \\
    $\Psi_{e}$    & 310.984 & 16.27457 & 0.00001
                            & $0.64(\Phi_1 - i \Phi_2)|S=7/2, M_s=5/2 \rangle + 0.26 i (\Phi_3 + i \Phi_4)|S=7/2, M_s=+7/2 \rangle$ \\
                  &         &          &
                            & $-0.64 i(\Phi_1 + i \Phi_2)|S=7/2, M_s=-5/2 \rangle + 0.26 (\Phi_3 - i \Phi_4)|S=7/2, M_s=-7/2 \rangle$ \\
    \hline
\end{tabular*}
\end{table*}

\subsection{Magnetic hyperfine interaction}

The magnetic hyperfine interaction originates from three microscopic interactions:\cite{AbragamBook,Sharkas2015} (i) the aforementioned FC contribution that represents the contact interaction of the electronic spin density at the nucleus with the nuclear spin; (ii) paramagnetic spin-orbital (PSO) term that describes coupling of electronic orbital angular momentum with the nuclear spin, and (iii) the spin-dipole (SD) terms that represents interaction between the electronic and nuclear spins.

We calculate the magnetic hyperfine interaction projected onto the ground Kramers doublet using the implementation and procedure discussed in Ref.~\citenum{Wysocki2020}. Note that the excited Kramers doublets are irrelevant to this hyperfine interaction because the excitation energy of the first-excited Kramers doublet (Fig.~\ref{fig:CAS}(c)) is at least two orders of magnitude higher than the level splitting of the nuclear spin levels. The {\it ab-initio} calculated electronic-nuclear energy spectrum is projected onto a model Hamiltonian with effective spin $S_{\rm{eff}}=1/2$ and the $^{159}$Tb nuclear spin ($I=3/2$) such as
\begin{eqnarray}
\hat{H}_\text{HF} &=& \hat{\mathbf{I}}\cdot\mathbf{A}\cdot\hat{\mathbf{S}}_{\rm{eff}},
\label{eq:H-th}
\end{eqnarray}
where $\mathbf{A}$ is the magnetic hyperfine matrix for the ground Kramers doublet.

\begin{table*}[h]
\small
\caption{\ The calculated hyperfine tensor elements (in units of MHz) for the ground Kramers doublet with and without external ${\bf E}$ field for the neutral Tb(II)(Cp$^{\rm{iPr5}}$)$_2$ molecule. Here the axes are chosen such that the ${\bf g}$ matrix in the absence of ${\bf E}$ field is diagonal. Here $|A_1|=\sqrt{A_{xz}^2+A_{yz}^2}/2$. The ${\bf E}$ field is applied along the $z$ axis}
\label{tab:hyperfine}
\begin{tabular*}{0.98\textwidth}{@{\extracolsep{\fill}}llllllll}
    \hline
    ${\bf E}$ field (mV/nm) & $A_{xx}$ & $A_{yy}$ & $A_{zz}$ & $A_{xy}$ & $A_{xz}$ & $A_{yz}$ & $|A_1|$ \\
    \hline
    0       &  0.00 & 0.00 & 42276.80 & 0.00 & $-$7.90  &   10.26 & 6.47  \\
    0.51    &  0.00 & 0.00 & 42263.50 & 0.00 & $-$10.56 &    3.10 & 5.50  \\
    5.14    &  0.00 & 0.00 & 42242.44 & 0.00 & $-$11.30 & $-$1.80 & 5.72  \\
    \hline
\end{tabular*}
\end{table*}

Table~\ref{tab:hyperfine} lists the ${\bf A}$ matrix elements with and without an ${\bf E}$ field, using the magnetic coordinates that diagonalize the ${\bf g}$ matrix. The largest element is $A_{zz}$ which is 42277~MHz in the absence of ${\bf E}$. This value is about seven times larger than that for the Tb(III)Pc$_2$ SMMs.\cite{Wysocki2020} The $A_{xx}$, $A_{yy}$, and $A_{xy}$ elements are very small such as an order of 10$^{-3}$ MHz because of the strong uniaxial magnetic anisotropy. The $A_{xz}$ and $A_{yz}$ elements are only about 0.02\% of the $A_{zz}$ element. The presence of nonzero $A_{xz}$ and $A_{yz}$ elements signals the deviation between the magnetic axes and the axes that diagonalize the $\mathbf{A}$ matrix. For isolated electronic $J$-multiplet, both sets of axes are expected to be identical.\cite{AbragamBook} For the neutral Tb(II)(Cp$^{\rm{iPr5}}$)$_2$, however, the low-energy electronic spectrum is a result of strong coupling between the $J=6$ multiplet and an extra electron occupying the Tb 5$d$/6$s$ orbitals. This nontrivial electronic structure is, thus, responsible for sizeable off-diagonal terms of the ${\bf A}$ matrix. The degree of the misalignment can be also estimated by introducing a new parameter such as $|A_1|=\sqrt{A_{xz}^2+A_{yz}^2}/2$.

\begin{figure}[h]
\centering
\includegraphics[height=4.5cm]{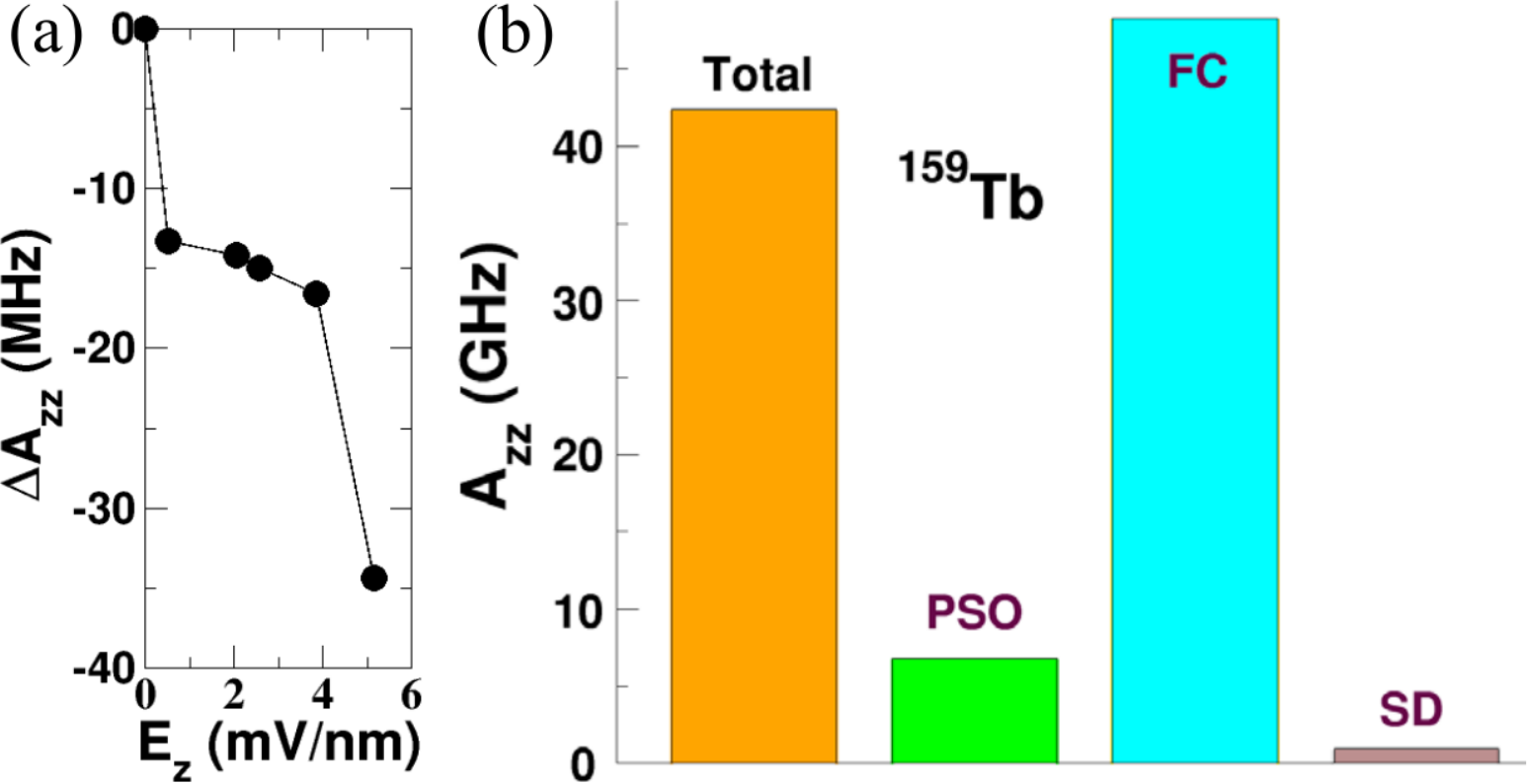}
\caption{(a) The change of the $A_{zz}$ element (in MHz) as a function of $E_z$ field relative to the zero-$E$ field value. (b)The contributions of the PSO, FC, and SD to the total hyperfine coupling parameter $A_{zz}$ for $^{159}$Tb(II)(Cp$^{\rm{iPr5}}$)$_2$ without ${\bf E}$ field. Note that the vertical scale is in GHz.}
\label{fig:hyperfine}
\end{figure}

Now when the $E$ field is applied along the $z$ axis, the $A_{zz}$ element substantially decreases with only small changes in the $A_{xz}$ and $A_{yz}$ elements or $|A_1|$, as shown in Table~\ref{tab:hyperfine}. Figure~\ref{fig:hyperfine}(a) shows the change of the $A_{zz}$ element as a function of ${\bf E}$ field using the magnetic axes without ${\bf E}$ field. The $A_{zz}$ element changes of an order of tens MHz with an ${\bf E}$ field of several mV/nm. The decrease in the $A_{zz}$ element is reflected in the change in the electronic-nuclear spectrum with ${\bf E}$ field that is discussed in Sec.~3.4.

In order to understand the nature of the hyperfine interaction and its $E$-field dependence, we calculate contributions of the PSO, FC, and SD terms to the $A_{zz}$ element. Figure~\ref{fig:hyperfine}(b) shows their contributions in the absence of ${\bf E}$ field. The FC term is dominant and it is about seven times larger than the PSO contribution. The large FC term is induced by the contribution of $(6s,5d_{z^2})$ orbital to the ground doublet (Tables~\ref{tab:SF} and \ref{tab:rassi}) which has a large weight at the nucleus. The $E$-field response to the $A_{zz}$ element is attributed to the change of the FC term. The nature of the hyperfine interaction in this molecule qualitatively differs from the case of Tb(III)Pc$_2$ SMMs where the PSO contribution is dominant and the FC contribution is negligible.\cite{Wysocki2020} Since the PSO contribution to the $A_{zz}$ element in the neutral Tb(II)(Cp$^{\rm{iPr5}}$)$_2$ is similar to that in the Tb(III)Pc$_2$ molecule, the total $A_{zz}$ element is about seven times larger than that in the Tb(III)Pc$_2$ molecule.

\subsection{Nuclear quadrupole interaction}

\begin{table*}[h]
\small
\caption{\ The calculated nuclear quadruple tensor elements (in units of MHz) for the ground Kramers doublet with and without external ${\bf E}$ field for the neutral Tb(II)(Cp$^{\rm{iPr5}}$)$_2$. Here the same coordinates as Table~\ref{tab:hyperfine} are used. The definitions of the additional parameters are as follows: $|P_1|=\sqrt{P_{xz}^2+P_{yz}^2}$ and $|P_2|=\sqrt{0.25(P_{xx}-P_{yy})^2+P_{xy}^2}$. The ${\bf E}$ field is applied along the $z$ axis}
\label{tab:quad}
\begin{tabular*}{0.98\textwidth}{@{\extracolsep{\fill}}lllllllll}
    \hline
    ${\bf E}$ field (mV/nm) & $P_{xx}$ & $P_{yy}$ & $P_{zz}$ & $P_{xy}$ & $P_{xz}$ & $P_{yz}$ & $|P_1|$ & $|P_2|$ \\
    \hline
    0       & $-$22.48 & $-$19.85 & 42.34 & 0.03 & $-$0.09 & 1.37 & 1.37 & 1.32  \\
    0.51    & $-$22.49 & $-$19.84 & 42.33 & 0.03 & $-$0.10 & 1.28 & 1.28 & 1.33  \\
    5.14    & $-$22.46 & $-$19.83 & 42.29 & 0.03 & $-$0.10 & 1.22 & 1.22 & 1.32  \\
    \hline
\end{tabular*}
\end{table*}

The nuclear quadrupole interaction is given by the following Hamiltonian:
\begin{equation}
\hat{H}_{\rm{quad}} = \hat{\mathbf{I}}\cdot\mathbf{P}\cdot\hat{\mathbf{I}},
\end{equation}
where ${\bf P}$ is nuclear quadrupole tensor projected onto the ground doublet of the neutral Tb(II)(Cp$^{\rm{iPr5}}$)$_2$. Table~\ref{tab:quad} lists the calculated ${\bf P}$ tensor with and without ${\bf E}$ field. The diagonal elements of the ${\bf P}$ tensor are approximately an order of magnitude smaller than those for Tb(III)Pc$_2$ SMMs.\cite{Wysocki2020} This feature suggests that the electric-field gradient at the Tb nuclear is quite small. The off-diagonal elements of the ${\bf P}$ tensor are very small like at most about 1 MHz because of the nearly-perfect symmetry of the neutral Tb(II)(Cp$^{\rm{iPr5}}$)$_2$ molecule. Compared to the $A_{zz}$ element, the response of the diagonal ${\bf P}$ elements to the $E_z$ field is three orders of magnitude smaller.

\subsection{Effective Spin Hamiltonian Analysis}

Based on our multiconfigurational calculations we construct an effective spin Hamiltonian that describes the low-energy electronic-nuclear spectrum of the neutral Tb(II)(Cp$^{\rm{iPr5}}$)$_2$ molecule. We focus on the electronic ground Kramers doublet and represent it by a pseudospin $S_{\rm{eff}}=1/2$. The low-lying nuclear levels are characterized by the Tb nucleus spin $I=3/2$. The effective Hamiltonian describing these two interacting system is given by
\begin{eqnarray}
\hat{H}_\text{eff}=\hat{H}_{\rm{HF}} + \hat{H}_{\rm{quad}} + \mu_{\text B}Bg_{zz}\hat{S}_{\text{eff}}^z+\mu_{\text N}Bg_\text{N}\hat{I}_z,
\label{Heff}
\end{eqnarray}
where the third and fourth terms describe the Zeeman interaction of, respectively, electronic and nuclear systems with the external ${\bf B}$ field, when the ${\bf B}$ field is applied along the $z$ axis. Here $g_\text{N}$ is the nuclear $g$-factor that for the $^{159}$Tb nucleus is 1.34267. As discussed in the previous subsections, $\mathbf{A}$, $\mathbf{P}$ and $g_{zz}$ are calculated from \emph{ab initio} and are shown in Tables~\ref{tab:rassi}, \ref{tab:hyperfine}, and \ref{tab:quad}.

\begin{figure}[t!]
\centering
\includegraphics[width=1.0\linewidth]{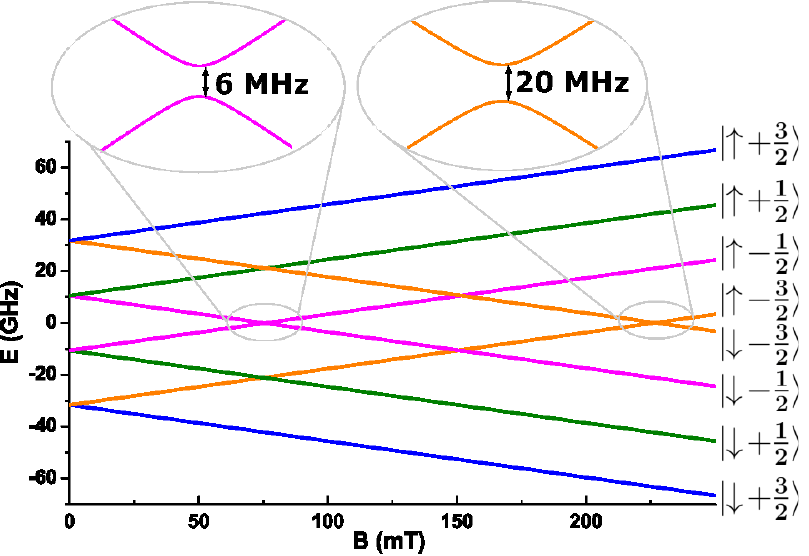}
\caption{Low-energy electronic-nuclear spectrum of the neutral Tb(II)(Cp$^{\rm{iPr5}}$)$_2$ molecule and its dependence on the external ${\bf B}$ field. Lines with positive (negative) slope correspond to levels with approximate $m_S$ quantum number equal to $\uparrow$ ($\downarrow$). The blue, green, magenta, and orange colors correspond to levels with approximate $M_I$ quantum number equal to $3/2$, $1/2$, $-1/2$, and $-3/2$, respectively. Zoom-in plots of crossing points between levels with the same approximate $M_I$ (top of the figure) show a significant avoided level crossing gap.}
\label{ZeemanDiagram}
\end{figure}

\subsubsection{Electronic-nuclear energy spectrum at zero magnetic field}

The electronic-nuclear spectrum obtained by diagonalization of the Hamiltonian~, Eq.~(\ref{Heff}), is shown in Fig.~\ref{ZeemanDiagram}. At zero magnetic field, the spectrum is composed of four doublets (the first- and second-excited doublets show a tiny splitting of the order of 10$^{-4}$ MHz due to presence of small but nonzero $A_{xx}$ and $A_{yy}$ parameters\cite{Wysocki2020b}). The levels can be characterized by $z$-projections of the electronic pseudospin ($m_S=\uparrow,\downarrow$) and the nuclear spin ($M_I=\pm3/2,\pm1/2$). Since the hyperfine interaction is the most dominant term in Eq.~(\ref{Heff}), the ground doublet corresponds to $|M_I|=3/2$ with $m_S$ being opposite sign to $M_I$. Further, the first-, second-, and third-excited doublets correspond to $\ket{\uparrow(\downarrow);-1/2(1/2)}$, $\ket{\uparrow(\downarrow);1/2(-1/2)}$, and $\ket{\uparrow(\downarrow);3/2(-3/2)}$ states, respectively. Note that $m_S$ and $M_I$ are not exactly good quantum numbers since transverse nuclear quadrupole terms and off-diagonal hyperfine interactions allow mixing between different $\ket{m_S,M_I}$ states. As shown in Tables~\ref{tab:rassi}, \ref{tab:hyperfine}, and \ref{tab:quad}, however, these interaction parameters are small due to fairly symmetric geometry of the neutral Tb(II)(Cp$^{\rm{iPr5}}$)$_2$ molecule and as a result, the mixing is small.

The separation between adjacent electronic-nuclear levels is about 21 GHz. This is significantly larger than for the Tb(III)Pc$_2$ molecules in which the level spacing is roughly 3 GHz.\cite{Wysocki2020} This difference is a direct result of a strong FC contribution for the neutral Tb(II)(Cp$^{\rm{iPr5}}$)$_2$ molecule which enhances the strength of the hyperfine interaction by almost an order of magnitude as compared to the Tb(III)Pc$_2$ case. Due to the quadrupole interaction (specifically $P_{zz}$ term), the levels are not equidistant and the level spacing increases with energy. Although the $P_{zz}$ term for the neutral Tb(II)(Cp$^{\rm{iPr5}}$)$_2$ ($\sim$0.04 GHz) is significantly smaller than for Tb(III)Pc$_2$ ($\sim$0.3 GHz),\cite{Wysocki2020} the variation of the spacing between adjacent levels ($\sim$0.1 GHz) is sufficiently large to distinguish transitions between different pairs of levels by spectroscopy.

\subsubsection{Zeeman diagram}

Let us now consider the behavior of the energy levels under application of the external ${\bf B}$ field directed along the $z$ axis (Fig.~\ref{ZeemanDiagram}). The main effect of the magnetic field comes from the electronic Zeeman term which causes the levels to vary linearly with the strength of the magnetic field. The states with $m_S=\uparrow$ vary with a positive slope while the states with $m_S=\downarrow$ vary with a negative slope. As a result, the zero-field degeneracy is removed. At certain field values, the levels with opposite $m_S$ cross. In the proximity of such crossing points, the small off-diagonal terms of the Hamiltonian [Eq.~(\ref{Heff})] become important and lead to strong mixing of the two states participating in the crossing. As a result, the avoided level crossing (ALC) occurs with a finite gap between the two crossing levels (see the insets of Fig.~\ref{ZeemanDiagram}). The ALC gap quantifies the strength of the state mixing and determines the probability of the tunneling between the two crossing levels as the ${\bf B}$ field is varied across the ALC point. It is, therefore, an important parameter describing the low-temperature dynamics of the system.

The largest gap from the ALC occurs between levels with the same $M_I$ (20 MHz for $M_I=3/2$). Although the magnitude of the gap is similar to that for anionic Tb(III)Pc$_2$,\cite{Wysocki2020} its microscopic mechanism is different from the latter case. For Tb(III)Pc$_2$, the ALC gap is caused by the transverse crystal fields and so it strongly depend on molecular geometry.\cite{Wysocki2020} On the other hand, the neutral Tb(II)(Cp$^{\rm{iPr5}}$)$_2$ molecule is a Kramers system and, thus, the time-reversal symmetry does prevent the presence of such transverse crystal fields in the effective Hamiltonian. Instead, the ALC gap is due to significant off-diagonal elements of the hyperfine matrix ($A_{xz}$, $A_{yz}$). The unusual electronic structure discussed in Sec.~3.1 is responsible for the significant ALC gap between levels with the same $M_I$ value.

\subsubsection{Hyperfine Stark effect}

\begin{figure}[t!]
\centering
\includegraphics[height=4.5cm]{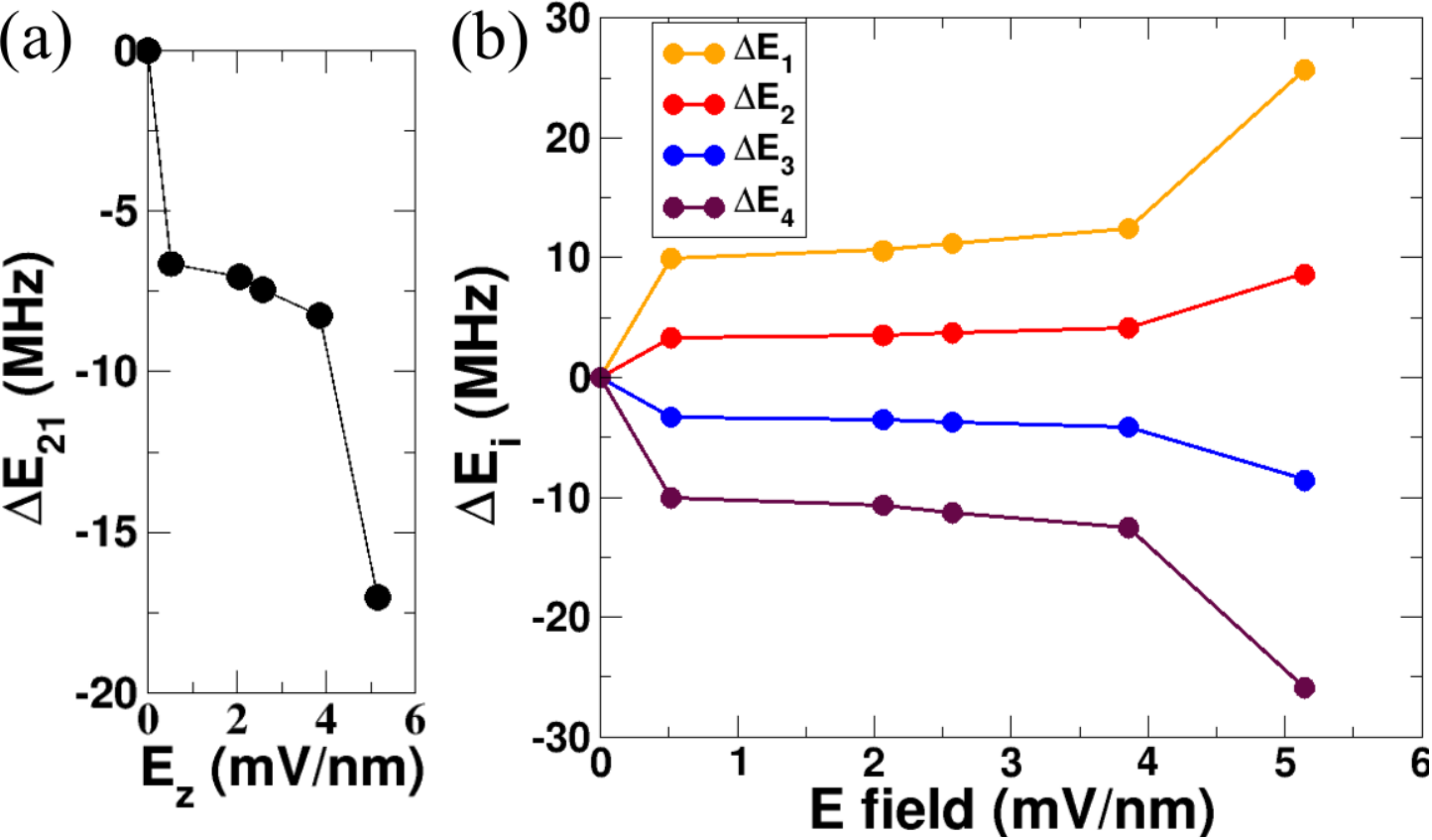}
\caption{(a) The excitation energy of the first-excited electronic-nuclear quasi-doublet vs ${\bf E}$ field. (b) The energy of each quasi-doublet relative to the zero-field energy ($\Delta E_i$, $i=1,2,3,4$) vs ${\bf E}$ field. From the top, the curves correspond to the ground, first-, second-, and third-excited quasi-doublet. In (a) and (b), the ${\bf E}$ field is applied along the $z$ axis and the ${\bf B}$ field is zero.}
\label{fig:EzDep}
\end{figure}

Let us now discuss how an external ${\bf E}$ field affects the electronic-nuclear spectrum when the field is applied along the $z$ axis. For simplicity, we consider only the case without an external ${\bf B}$ field. We calculate the energy eigenvalues of the effective Hamiltonian discussed earlier, Eq.~(\ref{Heff}), using the elements of the $\mathbf{A}$-matrix and the $\mathbf{P}$ tensor obtained with different values of the ${\bf E}$ field within the multiconfigurational approach. See Tables~\ref{tab:hyperfine} and \ref{tab:quad}. Note that the $A_{zz}$ parameter shows a significant dependence on the ${\bf E}$ field. This indicates that the electronic-nuclear levels can be electrically tuned. Figure~\ref{fig:EzDep}(a) shows the excitation energy of the first-excited electronic-nuclear quasi-doublet as a function of ${\bf E}$ field. With the field of 5 mV/nm, the excitation energy changes by about 17 MHz. In Fig.~\ref{fig:EzDep}(b) we plot the energy of each quasi-doublet at different ${\bf E}$ field values, relative to the zero-field energy of the quasi-doublet. This hyperfine Stark effect can be utilized in designing molecular qubits based on neutral Tb(II)(Cp$^{\rm{iPr5}}$)$_2$ molecules.

\section{Conclusions}

We have examined the electronic structure and the hyperfine and nuclear quadrupole interactions of the $^{159}$Tb nucleus in a divalent Tb compound, neutral Tb(II)(Cp$^{\rm{iPr5}}$)$_2$, using the CASSCF(13,14)-SO-RASSI method. Our calculation shows that the low-energy states arise from $4f^8(6s,5d_{z^2})^1$, $4f^8(5d_{x^2-y^2})^1$, and $4f^8(5d_{xy})^1$ configurations, where the spin from $4f^8$ is parallel to the spin from $(6s,5d)^1$. We found that the energy difference between the parallel and antiparallel spin configurations in $4f^8(6s,5d)^1$ is greater than the SO interaction and that the complexity of the electronic configuration demands a study of magnetic susceptibility beyond single-electron description. In addition, ${\bf J}$ is not a good quantum number due to the complex electronic configuration. 

The strong uniaxial magnetic anisotropy results in a large excitation energy of the electronic first-excited Kramers doublet (311~cm$^{-1}$), which is at least two orders of magnitude greater than the splitting of the electronic-nuclear levels for the Tb nuclear spin ($I=3/2$). Considering this large electronic excitation energy, we calculated the hyperfine and quadrupole interaction projected onto the electronic ground Kramers doublet using effective spin $S_{\rm{eff}}=1/2$ within CASSCF-SO-RASSI. We found that the FC contribution is dominant over PSO and SD contributions to the hyperfine interaction, because of the unusual electronic configuration of the Tb(II) ion. The hyperfine interaction for the neutral Tb(II)(Cp$^{\rm{iPr5}}$)$_2$ SMM turns out to be about one order of magnitude greater than that for Tb(III)Pc$_2$ SMMs. The dominant contribution of the FC term gives rise to tuning of the electronic-nuclear levels by tens of MHz with an electric field of an order of several mV/nm. Our findings stimulate future experiments on a search for the hyperfine Stark effect from divalent Tb compounds as well as other divalent lanthanide compounds and applications of the molecular nuclear spin levels for realization and operations of molecular spin qubits.

\section*{Conflicts of interest}
There are no conflicts to declare.

\section*{Acknowledgements}
This work was funded by the United States Department of Energy (DOE) Basic Energy Sciences (BES) grant number DE-SC0018326. Computational support by Virginia Tech ARC and San Diego Supercomputer Center (SDSC) grant number under DMR060009N.

\balance

\bibliography{refs} 

\providecommand*{\mcitethebibliography}{\thebibliography}
\csname @ifundefined\endcsname{endmcitethebibliography}
{\let\endmcitethebibliography\endthebibliography}{}
\begin{mcitethebibliography}{40}
\providecommand*{\natexlab}[1]{#1}
\providecommand*{\mciteSetBstSublistMode}[1]{}
\providecommand*{\mciteSetBstMaxWidthForm}[2]{}
\providecommand*{\mciteBstWouldAddEndPuncttrue}
  {\def\EndOfBibitem{\unskip.}}
\providecommand*{\mciteBstWouldAddEndPunctfalse}
  {\let\EndOfBibitem\relax}
\providecommand*{\mciteSetBstMidEndSepPunct}[3]{}
\providecommand*{\mciteSetBstSublistLabelBeginEnd}[3]{}
\providecommand*{\EndOfBibitem}{}
\mciteSetBstSublistMode{f}
\mciteSetBstMaxWidthForm{subitem}
{(\emph{\alph{mcitesubitemcount}})}
\mciteSetBstSublistLabelBeginEnd{\mcitemaxwidthsubitemform\space}
{\relax}{\relax}

\bibitem[Wernsdorfer and Ruben(2019)]{Wernsdorfer2019}
W.~Wernsdorfer and M.~Ruben, \emph{Adv. Mater.}, 2019, \textbf{31},
  1806687\relax
\mciteBstWouldAddEndPuncttrue
\mciteSetBstMidEndSepPunct{\mcitedefaultmidpunct}
{\mcitedefaultendpunct}{\mcitedefaultseppunct}\relax
\EndOfBibitem
\bibitem[Atzori \emph{et~al.}(2018)Atzori, Chiesa, Morra, Chiesa, Sorace,
  Carretta, and Sessoli]{Atzori2018}
M.~Atzori, A.~Chiesa, E.~Morra, M.~Chiesa, L.~Sorace, S.~Carretta and
  R.~Sessoli, \emph{Chem. Sci.}, 2018, \textbf{9}, 6183--6192\relax
\mciteBstWouldAddEndPuncttrue
\mciteSetBstMidEndSepPunct{\mcitedefaultmidpunct}
{\mcitedefaultendpunct}{\mcitedefaultseppunct}\relax
\EndOfBibitem
\bibitem[Gaita-Arino \emph{et~al.}({2019})Gaita-Arino, Luis, Hill, and
  Coronado]{Gaita2019}
A.~Gaita-Arino, F.~Luis, S.~Hill and E.~Coronado, \emph{{Nat. Chem.}}, {2019},
  \textbf{{11}}, {301--309}\relax
\mciteBstWouldAddEndPuncttrue
\mciteSetBstMidEndSepPunct{\mcitedefaultmidpunct}
{\mcitedefaultendpunct}{\mcitedefaultseppunct}\relax
\EndOfBibitem
\bibitem[Leuenberger and Loss(2001)]{Leuenberger2001}
M.~Leuenberger and D.~Loss, \emph{Nature}, 2001, \textbf{410}, 789--793\relax
\mciteBstWouldAddEndPuncttrue
\mciteSetBstMidEndSepPunct{\mcitedefaultmidpunct}
{\mcitedefaultendpunct}{\mcitedefaultseppunct}\relax
\EndOfBibitem
\bibitem[Awschalom \emph{et~al.}({2018})Awschalom, Hanson, Wrachtrup, and
  Zhou]{Awschalom2018}
D.~D. Awschalom, R.~Hanson, J.~Wrachtrup and B.~B. Zhou, \emph{{Nat.
  Photonics}}, {2018}, \textbf{{12}}, {516--527}\relax
\mciteBstWouldAddEndPuncttrue
\mciteSetBstMidEndSepPunct{\mcitedefaultmidpunct}
{\mcitedefaultendpunct}{\mcitedefaultseppunct}\relax
\EndOfBibitem
\bibitem[Thiele \emph{et~al.}(2014)Thiele, Balestro, Ballou, Klyatskaya, Ruben,
  and Wernsdorfer]{Thiele2014}
S.~Thiele, F.~Balestro, R.~Ballou, S.~Klyatskaya, M.~Ruben and W.~Wernsdorfer,
  \emph{Science}, 2014, \textbf{344}, 1135--1138\relax
\mciteBstWouldAddEndPuncttrue
\mciteSetBstMidEndSepPunct{\mcitedefaultmidpunct}
{\mcitedefaultendpunct}{\mcitedefaultseppunct}\relax
\EndOfBibitem
\bibitem[Shiddiq \emph{et~al.}(2016)Shiddiq, Komijani, Duan, Gaita-Ariño,
  Coronado, and Hill]{Shiddiq2016}
M.~Shiddiq, D.~Komijani, Y.~Duan, A.~Gaita-Ariño, E.~Coronado and S.~Hill,
  \emph{Nature}, 2016, \textbf{531}, 348--351\relax
\mciteBstWouldAddEndPuncttrue
\mciteSetBstMidEndSepPunct{\mcitedefaultmidpunct}
{\mcitedefaultendpunct}{\mcitedefaultseppunct}\relax
\EndOfBibitem
\bibitem[Godfrin \emph{et~al.}(2017)Godfrin, Ferhat, Ballou, Klyatskaya, Ruben,
  Wernsdorfer, and Balestro]{Godfrin2017}
C.~Godfrin, A.~Ferhat, R.~Ballou, S.~Klyatskaya, M.~Ruben, W.~Wernsdorfer and
  F.~Balestro, \emph{Phys. Rev. Lett.}, 2017, \textbf{119}, 187702\relax
\mciteBstWouldAddEndPuncttrue
\mciteSetBstMidEndSepPunct{\mcitedefaultmidpunct}
{\mcitedefaultendpunct}{\mcitedefaultseppunct}\relax
\EndOfBibitem
\bibitem[Moreno-Pineda \emph{et~al.}(2018)Moreno-Pineda, Klyatskaya, Du,
  Damjanović, Taran, Wernsdorfer, and Ruben]{Pineda2018}
E.~Moreno-Pineda, S.~Klyatskaya, P.~Du, M.~Damjanović, G.~Taran,
  W.~Wernsdorfer and M.~Ruben, \emph{Inorg. Chem.}, 2018, \textbf{57},
  9873--9879\relax
\mciteBstWouldAddEndPuncttrue
\mciteSetBstMidEndSepPunct{\mcitedefaultmidpunct}
{\mcitedefaultendpunct}{\mcitedefaultseppunct}\relax
\EndOfBibitem
\bibitem[Ishikawa \emph{et~al.}(2003)Ishikawa, Sugita, Ishikawa, Koshihara, and
  Kaizu]{Ishikawa2003}
N.~Ishikawa, M.~Sugita, T.~Ishikawa, S.-y. Koshihara and Y.~Kaizu, \emph{J. Am.
  Chem. Soc.}, 2003, \textbf{125}, 8694--8695\relax
\mciteBstWouldAddEndPuncttrue
\mciteSetBstMidEndSepPunct{\mcitedefaultmidpunct}
{\mcitedefaultendpunct}{\mcitedefaultseppunct}\relax
\EndOfBibitem
\bibitem[Thiele \emph{et~al.}(2013)Thiele, Vincent, Holzmann, Klyatskaya,
  Ruben, Balestro, and Wernsdorfer]{Thiele2013}
S.~Thiele, R.~Vincent, M.~Holzmann, S.~Klyatskaya, M.~Ruben, F.~Balestro and
  W.~Wernsdorfer, \emph{Phys. Rev. Lett.}, 2013, \textbf{111}, 037203\relax
\mciteBstWouldAddEndPuncttrue
\mciteSetBstMidEndSepPunct{\mcitedefaultmidpunct}
{\mcitedefaultendpunct}{\mcitedefaultseppunct}\relax
\EndOfBibitem
\bibitem[Ishikawa \emph{et~al.}(2005)Ishikawa, Sugita, and
  Wernsdorfer]{Ishikawa2005}
N.~Ishikawa, M.~Sugita and W.~Wernsdorfer, \emph{Angew. Chem.}, 2005,
  \textbf{44}, 2931--2935\relax
\mciteBstWouldAddEndPuncttrue
\mciteSetBstMidEndSepPunct{\mcitedefaultmidpunct}
{\mcitedefaultendpunct}{\mcitedefaultseppunct}\relax
\EndOfBibitem
\bibitem[Kane({1998})]{Kane1998}
B.~Kane, \emph{{Nature}}, {1998}, \textbf{{393}}, {133--137}\relax
\mciteBstWouldAddEndPuncttrue
\mciteSetBstMidEndSepPunct{\mcitedefaultmidpunct}
{\mcitedefaultendpunct}{\mcitedefaultseppunct}\relax
\EndOfBibitem
\bibitem[Gould \emph{et~al.}(2019)Gould, McClain, Yu, Groshens, Furche, Harvey,
  and Long]{Gould2019}
C.~A. Gould, K.~R. McClain, J.~M. Yu, T.~J. Groshens, F.~Furche, B.~G. Harvey
  and J.~R. Long, \emph{J. Am. Chem. Soc.}, 2019, \textbf{141},
  12967--12973\relax
\mciteBstWouldAddEndPuncttrue
\mciteSetBstMidEndSepPunct{\mcitedefaultmidpunct}
{\mcitedefaultendpunct}{\mcitedefaultseppunct}\relax
\EndOfBibitem
\bibitem[MacDonald \emph{et~al.}(2013)MacDonald, Bates, Ziller, Furche, and
  Evans]{MacDonald2013}
M.~R. MacDonald, J.~E. Bates, J.~W. Ziller, F.~Furche and W.~J. Evans, \emph{J.
  Am. Chem. Soc.}, 2013, \textbf{135}, 9857--9868\relax
\mciteBstWouldAddEndPuncttrue
\mciteSetBstMidEndSepPunct{\mcitedefaultmidpunct}
{\mcitedefaultendpunct}{\mcitedefaultseppunct}\relax
\EndOfBibitem
\bibitem[Woodruff \emph{et~al.}(2013)Woodruff, Winpenny, and
  Layfield]{Woodruff2013}
D.~N. Woodruff, R.~E.~P. Winpenny and R.~A. Layfield, \emph{Chem. Rev.}, 2013,
  \textbf{113}, 5110--5148\relax
\mciteBstWouldAddEndPuncttrue
\mciteSetBstMidEndSepPunct{\mcitedefaultmidpunct}
{\mcitedefaultendpunct}{\mcitedefaultseppunct}\relax
\EndOfBibitem
\bibitem[MacDonald \emph{et~al.}(2012)MacDonald, Bates, Fieser, Ziller, Furche,
  and Evans]{MacDonald2012}
M.~R. MacDonald, J.~E. Bates, M.~E. Fieser, J.~W. Ziller, F.~Furche and W.~J.
  Evans, \emph{J. Am. Chem. Soc.}, 2012, \textbf{134}, 8420--8423\relax
\mciteBstWouldAddEndPuncttrue
\mciteSetBstMidEndSepPunct{\mcitedefaultmidpunct}
{\mcitedefaultendpunct}{\mcitedefaultseppunct}\relax
\EndOfBibitem
\bibitem[Fieser \emph{et~al.}(2015)Fieser, MacDonald, Krull, Bates, Ziller,
  Furche, and E~vans]{Fieser2015}
M.~E. Fieser, M.~R. MacDonald, B.~T. Krull, J.~E. Bates, J.~W. Ziller,
  F.~Furche and W.~J. E~vans, \emph{J. Am. Chem. Soc.}, 2015, \textbf{137},
  369--382\relax
\mciteBstWouldAddEndPuncttrue
\mciteSetBstMidEndSepPunct{\mcitedefaultmidpunct}
{\mcitedefaultendpunct}{\mcitedefaultseppunct}\relax
\EndOfBibitem
\bibitem[Evans(2016)]{Evans2016}
W.~J. Evans, \emph{Organometallics}, 2016, \textbf{35}, 3088--3100\relax
\mciteBstWouldAddEndPuncttrue
\mciteSetBstMidEndSepPunct{\mcitedefaultmidpunct}
{\mcitedefaultendpunct}{\mcitedefaultseppunct}\relax
\EndOfBibitem
\bibitem[Fieser \emph{et~al.}(2017)Fieser, Palumbo, La~Pierre, Halter, Voora,
  Ziller, Furc~he, Meyer, and Evans]{Fieser2017}
M.~E. Fieser, C.~T. Palumbo, H.~S. La~Pierre, D.~P. Halter, V.~K. Voora, J.~W.
  Ziller, F.~Furc~he, K.~Meyer and W.~J. Evans, \emph{Chem. Sci.}, 2017,
  \textbf{8}, 7424--7433\relax
\mciteBstWouldAddEndPuncttrue
\mciteSetBstMidEndSepPunct{\mcitedefaultmidpunct}
{\mcitedefaultendpunct}{\mcitedefaultseppunct}\relax
\EndOfBibitem
\bibitem[Huh \emph{et~al.}(2018)Huh, Darago, Ziller, and Evans]{Huh2018}
D.~N. Huh, L.~E. Darago, J.~W. Ziller and W.~J. Evans, \emph{Inorg. Chem.},
  2018, \textbf{57}, 2096--2102\relax
\mciteBstWouldAddEndPuncttrue
\mciteSetBstMidEndSepPunct{\mcitedefaultmidpunct}
{\mcitedefaultendpunct}{\mcitedefaultseppunct}\relax
\EndOfBibitem
\bibitem[Ryan \emph{et~al.}({2018})Ryan, Darago, Balasubramani, Chen, Ziller,
  Furche, Long, and Evans]{Ryan2018}
A.~J. Ryan, L.~E. Darago, S.~G. Balasubramani, G.~P. Chen, J.~W. Ziller,
  F.~Furche, J.~R. Long and W.~J. Evans, \emph{{Chem. Eur. J.}}, {2018},
  \textbf{{24}}, {7702--7709}\relax
\mciteBstWouldAddEndPuncttrue
\mciteSetBstMidEndSepPunct{\mcitedefaultmidpunct}
{\mcitedefaultendpunct}{\mcitedefaultseppunct}\relax
\EndOfBibitem
\bibitem[Meihaus \emph{et~al.}(2015)Meihaus, Fieser, Corbey, Evans, and
  Long]{Meihaus2015}
K.~R. Meihaus, M.~E. Fieser, J.~F. Corbey, W.~J. Evans and J.~R. Long, \emph{J.
  Am. Chem. Soc.}, 2015, \textbf{137}, 9855--9860\relax
\mciteBstWouldAddEndPuncttrue
\mciteSetBstMidEndSepPunct{\mcitedefaultmidpunct}
{\mcitedefaultendpunct}{\mcitedefaultseppunct}\relax
\EndOfBibitem
\bibitem[Zhang \emph{et~al.}()Zhang, Muhtadi, Iwahara, Ungur, and
  Chibotaru]{Zhang2020}
W.~Zhang, A.~Muhtadi, N.~Iwahara, L.~Ungur and L.~F. Chibotaru, \emph{{Angew.
  Chem. Int. Ed.}}\relax
\mciteBstWouldAddEndPunctfalse
\mciteSetBstMidEndSepPunct{\mcitedefaultmidpunct}
{}{\mcitedefaultseppunct}\relax
\EndOfBibitem
\bibitem[Kova{\v{c}}evi{\'{c}} and Veryazov(2015)]{LUSCUS}
G.~Kova{\v{c}}evi{\'{c}} and V.~Veryazov, \emph{J. Cheminform.}, 2015,
  \textbf{7}, 16\relax
\mciteBstWouldAddEndPuncttrue
\mciteSetBstMidEndSepPunct{\mcitedefaultmidpunct}
{\mcitedefaultendpunct}{\mcitedefaultseppunct}\relax
\EndOfBibitem
\bibitem[Aquilante \emph{et~al.}(2016)Aquilante, Autschbach, Carlson,
  Chibotaru, Delcey, De~Vico, Fdez.~Galv{\'{a}}n, Ferr{\'{e}}, Frutos,
  Gagliardi, Garavelli, Giussani, Hoyer, Li~Manni, Lischka, Ma, Malmqvist,
  M{\"{u}}ller, Nenov, Olivucci, Pedersen, Peng, Plasser, Pritchard, Reiher,
  Rivalta, Schapiro, Segarra-Mart{\'{i}}, Stenrup, Truhlar, Ungur, Valentini,
  Vancoillie, Veryazov, Vysotskiy, Weingart, Zapata, and Lindh]{molcas}
F.~Aquilante, J.~Autschbach, R.~K. Carlson, L.~F. Chibotaru, M.~G. Delcey,
  L.~De~Vico, I.~Fdez.~Galv{\'{a}}n, N.~Ferr{\'{e}}, L.~M. Frutos,
  L.~Gagliardi, M.~Garavelli, A.~Giussani, C.~E. Hoyer, G.~Li~Manni,
  H.~Lischka, D.~Ma, P.-{\AA}. Malmqvist, T.~M{\"{u}}ller, A.~Nenov,
  M.~Olivucci, T.~B. Pedersen, D.~Peng, F.~Plasser, B.~Pritchard, M.~Reiher,
  I.~Rivalta, I.~Schapiro, J.~Segarra-Mart{\'{i}}, M.~Stenrup, D.~G. Truhlar,
  L.~Ungur, A.~Valentini, S.~Vancoillie, V.~Veryazov, V.~P. Vysotskiy,
  O.~Weingart, F.~Zapata and R.~Lindh, \emph{J. Comput. Chem.}, 2016,
  \textbf{37}, 506--541\relax
\mciteBstWouldAddEndPuncttrue
\mciteSetBstMidEndSepPunct{\mcitedefaultmidpunct}
{\mcitedefaultendpunct}{\mcitedefaultseppunct}\relax
\EndOfBibitem
\bibitem[Douglas and Kroll(1974)]{Douglass1974}
M.~Douglas and N.~M. Kroll, \emph{Ann. Phys.}, 1974, \textbf{82}, 89--155\relax
\mciteBstWouldAddEndPuncttrue
\mciteSetBstMidEndSepPunct{\mcitedefaultmidpunct}
{\mcitedefaultendpunct}{\mcitedefaultseppunct}\relax
\EndOfBibitem
\bibitem[Hess(1986)]{Hess1986}
B.~A. Hess, \emph{Phys. Rev. A}, 1986, \textbf{33}, 3742--3748\relax
\mciteBstWouldAddEndPuncttrue
\mciteSetBstMidEndSepPunct{\mcitedefaultmidpunct}
{\mcitedefaultendpunct}{\mcitedefaultseppunct}\relax
\EndOfBibitem
\bibitem[Roos \emph{et~al.}(1980)Roos, Taylor, and Siegbahn]{Roos1980}
B.~O. Roos, P.~R. Taylor and P.~E.~M. Siegbahn, \emph{Chem. Phys.}, 1980,
  \textbf{48}, 157--173\relax
\mciteBstWouldAddEndPuncttrue
\mciteSetBstMidEndSepPunct{\mcitedefaultmidpunct}
{\mcitedefaultendpunct}{\mcitedefaultseppunct}\relax
\EndOfBibitem
\bibitem[Siegbahn \emph{et~al.}(1981)Siegbahn, Alml{\"{o}}f, Heiberg, and
  Roos]{Siegbahn1981}
P.~E.~M. Siegbahn, J.~Alml{\"{o}}f, A.~Heiberg and B.~O. Roos, \emph{J. Chem.
  Phys.}, 1981, \textbf{74}, 2384--2396\relax
\mciteBstWouldAddEndPuncttrue
\mciteSetBstMidEndSepPunct{\mcitedefaultmidpunct}
{\mcitedefaultendpunct}{\mcitedefaultseppunct}\relax
\EndOfBibitem
\bibitem[Hess \emph{et~al.}(1996)Hess, Marian, Wahlgren, and Gropen]{Hess1996}
B.~A. Hess, C.~M. Marian, U.~Wahlgren and O.~Gropen, \emph{Chem. Phys. Lett.},
  1996, \textbf{251}, 365 -- 371\relax
\mciteBstWouldAddEndPuncttrue
\mciteSetBstMidEndSepPunct{\mcitedefaultmidpunct}
{\mcitedefaultendpunct}{\mcitedefaultseppunct}\relax
\EndOfBibitem
\bibitem[Malmqvist \emph{et~al.}(2002)Malmqvist, Roos, and
  Schimmelpfennig]{rassi}
P.-{\AA}. Malmqvist, B.~O. Roos and B.~Schimmelpfennig, \emph{Chem. Phys.
  Lett.}, 2002, \textbf{357}, 230--240\relax
\mciteBstWouldAddEndPuncttrue
\mciteSetBstMidEndSepPunct{\mcitedefaultmidpunct}
{\mcitedefaultendpunct}{\mcitedefaultseppunct}\relax
\EndOfBibitem
\bibitem[Nihei \emph{et~al.}(2007)Nihei, Shiga, Maeda, and Oshio]{Nihei2007}
M.~Nihei, T.~Shiga, Y.~Maeda and H.~Oshio, \emph{Coord. Chem. Rev.}, 2007,
  \textbf{251}, 2606 -- 2621\relax
\mciteBstWouldAddEndPuncttrue
\mciteSetBstMidEndSepPunct{\mcitedefaultmidpunct}
{\mcitedefaultendpunct}{\mcitedefaultseppunct}\relax
\EndOfBibitem
\bibitem[Halcrow(2011)]{Halcrow2011}
M.~A. Halcrow, \emph{Chem. Soc. Rev.}, 2011, \textbf{40}, 4119--4142\relax
\mciteBstWouldAddEndPuncttrue
\mciteSetBstMidEndSepPunct{\mcitedefaultmidpunct}
{\mcitedefaultendpunct}{\mcitedefaultseppunct}\relax
\EndOfBibitem
\bibitem[Wysocki and Park(2020)]{Wysocki2020}
A.~L. Wysocki and K.~Park, \emph{Inorg. Chem.}, 2020, \textbf{59},
  2771--2780\relax
\mciteBstWouldAddEndPuncttrue
\mciteSetBstMidEndSepPunct{\mcitedefaultmidpunct}
{\mcitedefaultendpunct}{\mcitedefaultseppunct}\relax
\EndOfBibitem
\bibitem[Chibotaru and Ungur(2012)]{chibotaru2012ab}
L.~F. Chibotaru and L.~Ungur, \emph{J. Chem. Phys.}, 2012, \textbf{137},
  064112\relax
\mciteBstWouldAddEndPuncttrue
\mciteSetBstMidEndSepPunct{\mcitedefaultmidpunct}
{\mcitedefaultendpunct}{\mcitedefaultseppunct}\relax
\EndOfBibitem
\bibitem[Anderson \emph{et~al.}(1989)Anderson, Cloke, Cox, Edelstein, Green,
  Pang, Sameh, and Shalimoff]{Anderson1989}
D.~M. Anderson, F.~G.~N. Cloke, P.~A. Cox, N.~Edelstein, J.~C. Green, T.~Pang,
  A.~A. Sameh and G.~Shalimoff, \emph{J. Chem. Soc.{,} Chem. Commun.}, 1989,
  53--55\relax
\mciteBstWouldAddEndPuncttrue
\mciteSetBstMidEndSepPunct{\mcitedefaultmidpunct}
{\mcitedefaultendpunct}{\mcitedefaultseppunct}\relax
\EndOfBibitem
\bibitem[Abragam and Bleaney(1970)]{AbragamBook}
A.~Abragam and B.~Bleaney, \emph{Electron Paramagnetic Resonance of Transition
  Ions}, Clarendon Press, Oxford, 1970\relax
\mciteBstWouldAddEndPuncttrue
\mciteSetBstMidEndSepPunct{\mcitedefaultmidpunct}
{\mcitedefaultendpunct}{\mcitedefaultseppunct}\relax
\EndOfBibitem
\bibitem[Sharkas \emph{et~al.}(2015)Sharkas, Pritchard, and
  Autschbach]{Sharkas2015}
K.~Sharkas, B.~Pritchard and J.~Autschbach, \emph{J. Chem. Theory Comput.},
  2015, \textbf{11}, 538--549\relax
\mciteBstWouldAddEndPuncttrue
\mciteSetBstMidEndSepPunct{\mcitedefaultmidpunct}
{\mcitedefaultendpunct}{\mcitedefaultseppunct}\relax
\EndOfBibitem
\bibitem[Wysocki and Park(2020)]{Wysocki2020b}
A.~L. Wysocki and K.~Park, \emph{J. Phys. Condens. Matter}, 2020, \textbf{32},
  274002\relax
\mciteBstWouldAddEndPuncttrue
\mciteSetBstMidEndSepPunct{\mcitedefaultmidpunct}
{\mcitedefaultendpunct}{\mcitedefaultseppunct}\relax
\EndOfBibitem
\end{mcitethebibliography}
\bibliographystyle{rsc} 

\end{document}